\documentclass[graybox]{svmult}

\usepackage{mathptmx}        
\usepackage{helvet}          
\usepackage{courier}         
\usepackage{type1cm}         

\usepackage{makeidx}          
\usepackage{graphicx}         
\usepackage{multicol}         
\usepackage[bottom]{footmisc} 

\usepackage{url}

\makeindex                    

\begin{document}

\title{New applications of Graph Neural Networks in Cosmology}

\author{Farida Farsian, Federico Marulli, Lauro Moscardini, Carlo Giocoli}
\authorrunning{Farsian et al.}

\institute{Farida Farsian \at Dipartimento di Fisica e Astronomia ``Augusto Righi'' - Alma Mater Studiorum Universit\`{a} di Bologna, via Piero Gobetti 93/2, I-40129 Bologna, Italy \\ INAF -
  Osservatorio di Astrofisica e Scienza dello Spazio di Bologna, via
  Piero Gobetti 93/3, I-40129 Bologna, Italy
  \email{farida.farsian@unibo.it} \and Federico Marulli \at Dipartimento di Fisica e Astronomia ``Augusto Righi'' - Alma Mater
  Studiorum Universit\`{a} di Bologna, via Piero Gobetti 93/2, I-40129
  Bologna, Italy \\ INAF - Osservatorio di Astrofisica e Scienza dello
  Spazio di Bologna, via Piero Gobetti 93/3, I-40129 Bologna, Italy
  \\ INFN - Sezione di Bologna, viale Berti Pichat 6/2, I-40127
  Bologna, Italy \email{federico.marulli3@unibo.it} \and Lauro
  Moscardini \at Dipartimento di Fisica e Astronomia ``Augusto Righi'' - Alma Mater Studiorum Universit\`{a} di Bologna, via Piero Gobetti 93/2, I-40129 Bologna, Italy \\ INAF - Osservatorio di Astrofisica e Scienza dello Spazio di Bologna, via Piero Gobetti 93/3, I-40129 Bologna, Italy \\ INFN - Sezione di Bologna, viale Berti Pichat 6/2, I-40127 Bologna, Italy \email{lauro.moscardini@unibo.it} \and Carlo Giocoli \at INAF - Osservatorio di Astrofisica e Scienza dello
  Spazio di Bologna, via Piero Gobetti 93/3, I-40129 Bologna,
  Italy \\ INFN - Sezione di Bologna, viale Berti Pichat 6/2, I-40127
  Bologna, Italy \email{carlo.giocoli@inaf.it}}

\maketitle

\abstract{Upcoming cosmological surveys will provide unprecedented
  amount of data, which will require innovative statistical methods to
  maximize the scientific exploitation. Standard cosmological analyses
  based on abundances, two-point and higher-order statistics of cosmic
  tracers have been widely used to investigate the properties of the
  {\em cosmic web} and Large Scale Structure. However, these statistics can only exploit a subset of the entire information content available. Our goal is thus to implement new data analysis techniques based on machine learning to extract cosmological information through forward
  modelling, by directly exploiting the spatial coordinates and other
  observed properties of galaxies and galaxy clusters. Specifically,
  we investigated a new representation of large-scale structure data
  in the form of {\em graphs}. This data format can be directly fed to
  {\em Graph Neural Networks}, a recently proposed class of supervised
  Deep Learning algorithms. We tested the method on dark matter halo
  catalogues in different cosmologies, finding promising results. In
  particular, the method can discriminate among different dark energy
  models with high accuracy, through both binary classification
  ($99\%-$accuracy) and multi-class classification
  ($97\%-$accuracy). Moreover, it provides constraints on the value of
  $w_0$, through regression, with high precision.}

\section{Introduction}
\label{intro}
Accurate and fast statistical data analysis techniques are essential
to extract the full cosmological information content in the data
provided by current and next-generation cosmological experiments
(e.g. Euclid \cite{Euclid} and LSST). A possible response to these needs could
be pulling the information out of raw data available in galaxy or
galaxy cluster catalogues. To do so, we propose to consider the
large-scale structure data in the form of graphs and let a so-called
Graph Neural Network (GNN) to obtain all the specifications of the
tracer distribution field.

GNNs are a class of Deep Learning methods designed to perform
inferences on irregular and sparse data described by graphs
\cite{general_gnn}. Indeed, they can capture the graph structure of
data which is often very rich, and are particularly suitable for
learning global permutation invariant quantities, which makes them
ideal to be used with large-scale galaxy catalogues \cite{Zhou et
  al.}.

\section{Graph Neural Networks}
\label{GNN}

To build a GNN for cosmological inference analyses, we have used the
Spektral package\footnote{https://graphneural.network}, which is based
on \texttt{Tensorflow} and \texttt{Keras}.  Our GNN architecture
consists of two blocks of EdgeGNN \cite{EdgeGNN} and two blocks of
GeneralConv \cite{GeneralConv}, followed by the pooling layers --
TopKpool layers \cite{Graph_Unet} and the Global average pooling
layer.

We test our GNN model on a large set of dark matter halo catalogues
extracted from the Quijote N-body simulations \cite{Quijote}, which
provide enough statistics to train Neural Networks on cosmologies with
different dark energy equation of state parameters. The box size of this simulation is 1 $Gpc/h$ with more than 8.5 trillions of particles at a single redshift. In particular, we consider $500$ mock catalogues extracted from the Quijote simulations at redshift $z=0$, and for three $w_0$ values: $[-0.95, -1.0, -1.05]$. Massive halos are the most relevant probes for the analysis performed in this work, and we have applied a mass halo cut of $7\times 10^{14}\,\mbox{M}_\odot$ which means we have around one thousand halos per realization.

The first step of the implemented method consists in building the
graph assigned to each dark matter halo catalogue, to give it as input
to the GNN. Each halo is considered as a node of the graph, while the
halo mass and coordinates are held as node features. Then we assign an
edge between nodes. Two nodes, $i$ and $j$, have an edge ($e_{i
  \rightarrow j}$) if they are closer than a certain distance $r$. We
keep $r$ as one of the hyper-parameters of the method, and implemented
a grid search algorithm to optimise its value. The results presented
here are obtained with $r = 100 $ $Mpc/h$.

\section{Results}
\label{res}
To assess the performance of the implemented GNN on the built cosmic
graphs, we check its ability in distinguishing and classifying halo
graphs with different values of $w_0$. Then we move to the regression
problem, which is more desirable for the purpose of this project. For
all the cases, we performed a graph-level analysis on $360$
realizations as the training, $40$ as the validation and $100$ as the
test set.

Firstly, we perform a binary classification on $w_0=[-0.95,
  -1.05$]. The implemented GNN is able to distinguish among these
values with $100\%$ of accuracy in both the training and validation
phases, and $99\%$ accuracy for the test set. The next stage of our
analysis is dedicated to a multi-class classification, with the three
values of $w_0$ as classes. In this case, the GNN reaches $99\%$ of
accuracy in the training and validation phases, and $97\%$ accuracy on
the test set.

\begin{figure}[t]
\sidecaption[t]
\includegraphics[width=0.485\textwidth]{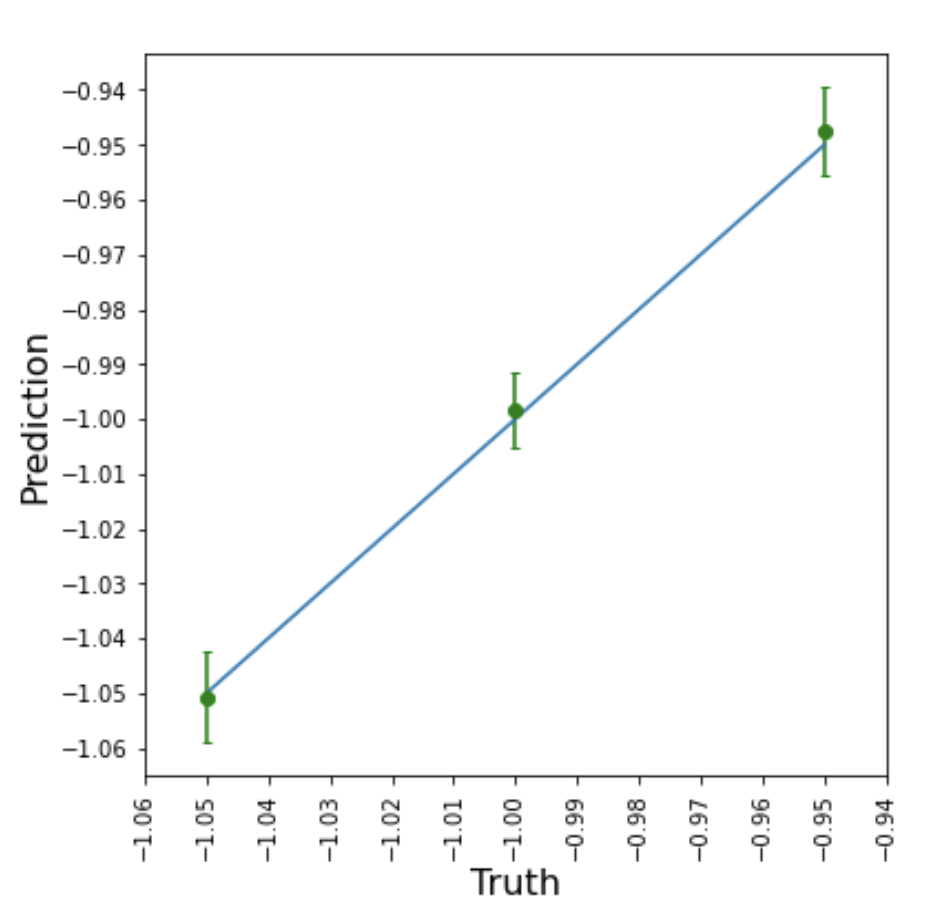}
\caption{Predictions of the implemented GNN on $w_0$. The true $w_0$
  value is shown on the y axis. The error bar indicates the dispersion
  of GNN prediction among the $100$ test set realizations.}
\label{fig1}
\end{figure}

As a final step, we exploit the GNN to provide a prediction on the
value of $w_0$ for the constructed halo graphs.  As it is shown in
Fig.  \ref{fig1}, the GNN predicts the value of $w_0$ very accurately,
with $2\%$ of statistical error for all the three sets of halo
catalogues considered. The error bars show the standard deviation of
the GNN predictions for the $100$ test set realizations.

%
%
%

\begin{thebibliography}{99.}%
\bibitem{Euclid}
A.~Blanchard \textit{et al.} [Euclid],
Astron. Astrophys. \textbf{642} (2020), A191
doi:10.1051/0004-6361/202038071
[arXiv:1910.09273 [astro-ph.CO]].

\bibitem{general_gnn} Wu Z., Pan S., Chen F., Long G., Zhang C., Yu P.~S., 2019, arXiv, arXiv:1901.00596.

\bibitem {Zhou et al.} Zhou J., Cui G., Hu S., Zhang Z., Yang C., Liu Z., Wang L., et al., 2018, ``Graph neural networks: A review of methods and applications'', arXiv, arXiv:1812.08434

\bibitem{EdgeGNN} Wang Y., Sun Y., Liu Z., Sarma S.~E., Bronstein M.~M., Solomon J.~M., 2018, ``Dynamic Graph CNN for Learning on Point Clouds'', arXiv, arXiv:1801.07829

\bibitem{GeneralConv} You J., Ying R., Leskovec J., 2020, ``Design Space for Graph Neural Networks'', arXiv, arXiv:2011.08843

\bibitem{Graph_Unet} Gao H., Ji S., 2019, ``Graph U-Nets'', arXiv, arXiv:1905.05178

\bibitem{Quijote}
F.~Villaescusa-Navarro, C.~Hahn, E.~Massara, A.~Banerjee, A.~M.~Delgado, D.~K.~Ramanah, T.~Charnock, E.~Giusarma, Y.~Li and E.~Allys, \textit{et al.}
``The Quijote simulations,''
Astrophys. J. Suppl. \textbf{250} (2020) no.1, 2
doi:10.3847/1538-4365/ab9d82
[arXiv:1909.05273 [astro-ph.CO]].





\end{thebibliography}

\end{document}